\documentclass[a4paper]{EslabStyle}
\usepackage{epsfig}

\begin{document}

\title{The Star Formation History of the Local Group}

\author{Eva K. Grebel\inst{1,2,3}} 
\institute{University of Washington, Astronomy Department, Box 351580,
  Seattle, WA 98195, USA \and Hubble Fellow \and Max-Planck-Institut
  f\"ur Astronomie, K\"onigstuhl 17, D-69117 Heidelberg, Germany}

\maketitle 

\begin{abstract}

The star formation histories of Local Group galaxies are summarized.
The three large spirals are discussed individually. The discussion
of the Local Group dwarfs concentrates on differences and commonalities.
While Local Group galaxies exhibit an amazing diversity in their 
star formation and enrichment histories even within the same morphological
type, they share a number of common global properties.  Major determining
factors for their evolution are galaxy mass and environmental effects  
such as ram pressure and tidal stripping.  It seems likely that low-mass
dwarf irregular galaxies evolve into dwarf spheroidals.

\end{abstract}

\section{Introduction}
\label{sec:Intro}

Our \object{Milky Way} and its companions are part of a sparse group of
galaxies known as the \object{Local Group} (LG).  Due to
the uncertainties in the mass estimates for LG galaxies and the lack of
information about galaxy orbits the LG boundaries are poorly defined.
The LG zero-velocity surface has a radius of $\sim 1.2$ Mpc (\cite{Court99}) 
around the LG barycenter when a simplified spherical potential is adopted. 
Within this zero-velocity
surface there are currently 36 galaxies known, most or all of which may be 
members of the LG.  This radius definition excludes galaxies such as
\object{SagDIG} or the nearby 
\object{NGC 3109}/Sextans 
minigroup (\cite{vdB99a}), which has a mean barycentric 
distance of 1.6 Mpc from the LG and is kinematically offset from the $\pm 60$
km s$^{-1}$ velocity dispersion envelope of the LG.  
In contrast to earlier definitions
(\cite{Grebel97}; \cite{Mateo98}) this radius also avoids
overlap with the closest members of the elongated \object{Sculptor group}, 
which are located at $\ga 1.7$ Mpc.  When viewed on a larger scale,
the Local Group appears to be one of several condensations in a clumpy 
filament that extends over many megaparsecs and contains a number of 
nearby groups and clusters (\cite{Tully87}).  

\section{Galaxy types in the Local Group}
\label{sec:types}

\begin{table*}[bht]
  \caption{A summary of global properties shared by many of the three main 
      dwarf galaxy types in the Local Group. ``R'' denotes typical tidal radii.
      ``SFR'' refers to the present-day star formation rate.  Individual dwarf
      galaxies may differ from the listed values.  For comparison the inferred
      properties of compact high velocity clouds (CHVCs) are also listed.
      }
  \label{tab1}
  \begin{center}
    \leavevmode
    \footnotesize
    \begin{tabular}[h]{lcccccr}
      \hline \\[-5pt]
Type & $M_V$ & $\mu_V$ & R & SFR & $M_{\rm H\,I}$ & $M_{tot}$\\%[+5pt]
& [mag] & [mag arcsec$^{-2}$] & [kpc] & [$M_{\odot}$ yr$^{-1}$] & 
[$M_{\odot}$] & [$M_{\odot}]$ \\[+5pt]
      \hline \\[-5pt]
dIrr & $\ga-18$ & $\la 23$ & $\la 5$ & $\la 0.01$ & $\la 10^9$ & $\la 10^{10}$ \\
dE & $\ga-17$ & $\la 21$ & $\la 4$ &      0     & $\la 10^8$ & $\la 10^9$ \\
dSph & $\ga-14$ & $\ga 22$ & $\la 3$ &      0     & $\la 10^5$ & $\sim 10^7$ \\
      \hline \\[-5pt]
CHVCs & ? & ? & $\la 10$ & 0? & $\la 10^7$ & $\ga 10^8$ \\
\hline \\
      \end{tabular}
  \end{center}
\end{table*}

The absolute magnitudes of LG
galaxies range from $\sim -21 < M_V < -8.5$ mag and comprise masses of
a few $10^{12} M_{\odot}$ to $10^7 M_{\odot}$.
As in any galaxy group and cluster
the majority of the galaxies in the LG are \object{dwarf galaxies}.  
If we consider
all galaxies with $M_V>-18$ mag to be dwarf galaxies, then the LG contains
31 dwarfs.  Three of the four galaxies that are more luminous than this
cutoff are spirals (the Milky Way, M31, \object{M33}), and one is a gas-rich, 
active irregular galaxy (the \object{Large Magellanic Cloud} (\object{LMC})).
The LMC is the closest 
irregular satellite of the Milky Way, and M33 is part of the M31 
subgroup. 

The LG dwarfs can be subdivided in three basic types: dwarf
irregulars (dIrrs), dwarf ellipticals (dEs), and dwarf spheroidals (dSphs).  
Note that these morphological distinctions are arbitrary to some
extent and are used in differing ways by different authors.  
In the following the term ``old population'' refers to ages $>10$ Gyr,
``intermediate age'' denotes ages from 1 Gyr to 10 Gyr,
and ``young'' includes all stellar populations below 1 Gyr.  

\subsection{Local Group dwarf irregulars}
\label{sec:dIrrs}

DIrrs tend to be gas-rich, irregularly shaped galaxies that show 
scattered, ongoing star formation while spiral density waves are absent.
Most dIrrs contain stellar populations of all ages as well as neutral and
molecular gas and dust.
Some of the LG irregulars/dIrrs show evidence for a bar (e.g., 
LMC, \object{NGC 6822}), while in the Small Magellanic Cloud (\object{SMC}) 
the apparent presence of a stellar bar may be feigned by highly
irregular recent star formation (\cite{Zar00}).

\subsection{Local Group dwarf ellipticals}
\label{sec:dEs}

DEs stand out through their distinctive elliptical shape and compact, dense 
cores.  Their dominant stellar populations are of old and intermediate age,
but recent star formation is observed as well.  As pointed out
by \cite*{vdB99b} and references therein, the structural parameters of
the M31 companion \object{M32} make this dwarf galaxy a genuine elliptical 
rather than a dE/dSph.  The remaining three dEs (\object{NGC
205}, \object{NGC 185}, and \object{NGC 147}; all M31 satellites) follow
similar scaling relations as the LG dSphs  (e.g., \cite{Ferg94}) and are
not always distinguished from dSphs in the literature.  Little to no gas 
(H\,{\sc i} and CO) has been detected in LG dEs (e.g., 
\cite{Sage98}).
Both M32 and NGC\,205 have a luminous semi-stellar core and may be 
considered nucleated (dE,N).  M32 appears to contain a massive central 
black hole (\cite{Marel97}).

\subsection{Local Group dwarf spheroidals}
\label{sec:dSphs}

\begin{figure}[htb]
  \begin{center}
    \epsfig{file=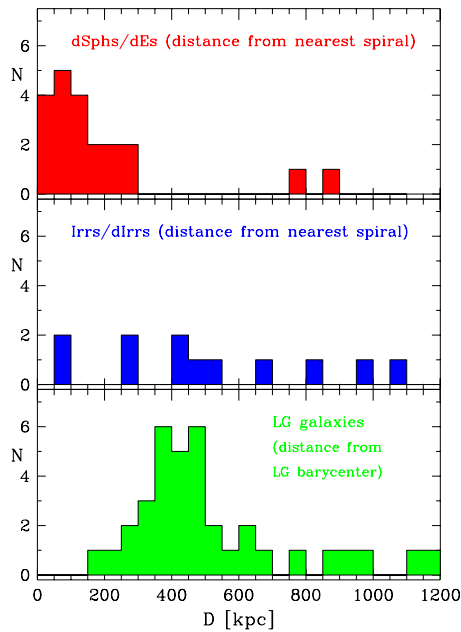, width=7cm}
  \end{center}
\caption{Number distribution of Local Group galaxies as a function of 
         distance.  Morphological segregation is evident.  
         Upper panel:\, Dwarf ellipticals and dwarf spheroidal galaxies 
         (the least massive dwarfs) are
         mostly found in close proximity of M31 or the Milky Way. 
         Middle panel:\, Dwarf irregular galaxies are more
         uniformly distributed throughout the Local Group.  The less
         massive galaxies among the dwarf irregulars with $250 < D_{Spiral} 
         < 450$ kpc are often referred to as transition-type objects (see text).
         Lower panel:\, Local Group galaxies tend to cluster around
         the two massive spirals.  Local Group membership for the outlying 
         objects at larger distances is uncertain. 
\label{fig1}}
\end{figure}

DSphs as defined by \cite*{Gall94} do not show nuclei and have very little
central concentration.  They are the least massive and least luminous
galaxies known.  Their stellar populations are predominantly of old and
intermediate age.  The lack of H\,{\sc i} at the optical positions of
these galaxies 
not understood.  DSphs, once thought to be an extension of globular clusters,
differ from globulars in several crucial ways: (1) they have very low stellar
densities even in their centers, (2) most of them have experienced multiple
or extended episodes of star formation, (3) they have high 
mass-to-light ratios (\cite{Mateo98}, but see \cite{Klessen98} for 
a different view), and (4) some dSphs contain globular clusters themselves.
Except for \object{Tucana} and
\object{Cetus}, LG dSphs are close companions of either the Milky Way or M31
(Fig.\ \ref{fig1}).  

A summary of representative
dwarf properties in the LG is given in Table \ref{tab1}. 
For further discussion of
morphological types in and around the Local Group we refer to
\cite*{Grebel99} and the very detailed reviews by van den Bergh (1999b, 2000a).
Globular clusters have been 
identified in all galaxy types in the LG, though only 13 LG 
galaxies contain globulars (see \cite{Grebel00} for a review).

\subsection{Completeness and missing satellites}
\label{sec:complete}

Many of the faint LG 
members were discovered only in recent years.
It is conceivable that the galaxy census continues to be
incomplete for faint 
dwarf galaxies and in regions of high foreground extinction.
Destruction of dwarfs through accretion by spirals may have played an
important role in the past and is currently evidenced by the merger of
the Sagittarius dSph galaxy with the Milky Way (\cite{Ibata94}).   
A comparison with the luminosity functions of nearby groups 
(\cite{Trent98})
suggests that the LG census is still deficient.  Deep wide-field surveys
such as the Digitized Second Palomar Sky Survey (\cite{Djor99}) and the 
Sloan Digital Sky Survey (\cite{Gunn95}) may lead to the detection of 
additional LG dwarfs of very low surface brightness.

Hierarchical cold dark matter (CDM) models predict roughly a factor
10 more dark matter halos for the LG than the number of known 
dwarf galaxies (\cite{Klypin99}; \cite{Moore99}).  If the CDM model predictions 
are correct and if these dark halos have baryonic counterparts, then
\object{compact high velocity clouds} (CHVCs; \cite{Braun99}) may be 
good candidates for the predicted LG satellites.  While most high-velocity
cloud complexes
are within 10 kpc of the Milky Way or part of the Magellanic stream,
CHVCs appears to be at  distances of up to 1 Mpc, show infall 
motion with respect to the LG barycenter,
have masses of $\sim 10^7 M_{\odot}$, are   
rotationally supported, and dark matter dominated (\cite{Braun00}).  
So far it is unknown whether CHVCs have stellar counterparts or might 
constitute low-mass protogalaxies prior to star formation.

\subsection{Morphological segregation}
\label{sec:segre}

The galaxy distribution in the LG is dominated by two prominent
subgroups around its two most massive members, the Milky Way and \object{M31}
(Fig.\ \ref{fig1}).  70\% of the smaller galaxies are found within $\sim
300$ kpc around the two dominant Sb/Sbc spirals.
\cite*{Wilk99} suggest that more than 50\% of the LG mass is concentrated 
around the two large spirals.
There are no massive ellipticals in the LG, and the third spiral, M33, does
not have any known companions.  The dependence of distribution within the
LG on galaxy type is also known as morphological segregation and may hold 
clues to the impact of environmental effects on the evolution of dwarf
galaxies in groups.  Morphological segregation is also evident in the nearby
M81 group (\cite{Kara00}). On a much larger scale similar effects may be seen 
in the morphology-density relation in galaxy clusters at low and high redshift
(\cite{Dressler80}; \cite{Dressler97}).

\begin{table*}[bht]
  \caption{Known and possible Local Group members within a barycentric radius
of 1.2 Mpc.  The galaxies are listed by subgroup (Milky Way and likely 
companions, M31 and likely companions, and galaxies at distances $> 500$ kpc
from one of the spirals) and sorted by increasing distance.
Galaxy types follow the nomenclature of \protect\cite*{vdB94} 
with addition of the letter ``N'' for Sgr, NGC\,205, and M32 to indicate that
these may be nucleated dwarfs.  $D_{\odot}$ denotes the distance from the 
Sun.  References are listed in the following column (``Ref'': 1 = 
\protect\cite{Bella99}, 2 = \protect\cite{Wester97}, 
3 = \protect\cite{Migh99}, 4 = \protect\cite{Grill98},
5 = \protect\cite{Kal95}, 6 = \protect\cite{Mateo95},
7 = \protect\cite{Migh97}, 8 = \protect\cite{Savi00},
9 = \protect\cite{Migh96}, 10 = \protect\cite{Lee93},
11 = \protect\cite{Held99}, 12 = \protect\cite{Gall96}, 
13 = \protect\cite{Freed90}, 14 = \protect\cite{Grill96},
15 = \protect\cite{Lee96}, 16 = \protect\cite{DaCosta00},
17 = \protect\cite{Armand93}, 18 = \protect\cite{Han97},
19 = \protect\cite{Armand98}, 20 = \protect\cite{Mart98},
21 = \protect\cite{Freed91}, 22 = \protect\cite{GrGu99},
23 = \protect\cite{Sakai99}, 24 = \protect\cite{Armand99},
25 = \protect\cite{Lee95}, 26 = \protect\cite{Gall98},
27 = \protect\cite{Cole99}, 28 = \protect\cite{Whit99},
29 = \protect\cite{Tols98}, 30 = \protect\cite{Dolp00},
31 = \protect\cite{Savi96},
32 = \protect\cite{Lee99}, 33 = \protect\cite{Lee00}).
$D_{\rm M31}$ from M31, and $D_{\rm LG}$ from the Local Group 
barycenter.  The barycenter was assumed to be located at
$l=121.7\degr$, $b=-21.3\degr$ (\protect\cite{Court99}), and 462 kpc.  
References for 
the absolute $V$ magnitudes can be found in \protect\cite*{vdB00b}.  Central 
surface brightnesses were taken from \protect\cite*{Mateo98} and complemented 
by data from \protect\cite*{Cald99} for And\,V, VI, and Cas\,dSph,
from \protect\cite*{Whit99} for Cetus, from \protect\cite*{Lee99} for DDO\,210, 
and \protect\cite*{Lauer98} for M33 and M31.  
The mean metallicities were taken 
from a variety of sources (column ``Ref''; 34 = \protect\cite{Shet98},
35 = \protect\cite{Mateo98}, 36 \protect\cite{Smec00},
37 = \protect\cite{GrSt99},  
38 = \protect\cite{Geis99}, 39 = \protect\cite{Guha00}, 
40 = \protect\cite{Cote99}, 41 = \protect\cite{Tikh99}, 42 = 
\protect\cite{Cook99}).
These values are in part based on red giant spectra or
on photometry and represent mean abundances of the dominant intermediate/old 
population.  Due to considerable enrichment and mixed populations
in some of the galaxies mean metallicities may not always a useful concept.  
The $\sigma$ quoted for $\rm\langle[Fe/H]\rangle$ indicates the metallicity
spread, not the uncertainty in the $\rm\langle[Fe/H]\rangle$ determination.  
      }
\label{tab2}
\begin{center}
    \leavevmode
    \footnotesize
    \begin{tabular}[h]{llrrrcccrccc}
      \hline \\[-5pt]
Name     & Type     & l~~~ & b~~~     & $D_{\odot}$~~ & Ref& $D_{\rm M31}$ & 
$D_{\rm LG}$ & $M_V$ & \protect$\mu_V$ & 
\protect$\rm\langle[Fe/H]\rangle$ & Ref \\ 
         &          & [\degr]~~ & [\degr]~~ & [kpc]~~ &    & [Mpc] & [Mpc] & [mag] & [mag arcsec$^{-2}$] & [dex] & \\
      \hline \\[-5pt]
Galaxy   &S(B)bcI-II&  0.00& $  0.00$ & $  8\pm~1$ &  &0.77 &0.47 &$-20.9$&  ---       & ---     &   \\
Sgr      & dSph,N?  &  6.00& $-15.00$ & $ 28\pm~3$ & 1&0.78 &0.47 &$-13.8$&$25.4\pm0.2$&$-1.0\pm0.3$ & 1 \\
LMC      & IrIII-IV &280.46& $-32.89$ & $ 50\pm~5$ & 2&0.80 &0.49 &$-18.5$&$20.7\pm0.1$&$-0.7\pm0.4$ & 2 \\
SMC      & IrIV/IV-V&302.80& $-44.30$ & $ 63\pm 10$& 2&0.80 &0.49 &$-17.1$&$22.1\pm0.1$&$-1.0\pm0.4$ & 2 \\
UMi      & dSph     &104.95& $ 44.80$ & $ 69\pm~4$ & 3&0.75 &0.44 &$ -8.9$&$25.5\pm0.5$&$-2.2\pm0.2$ & 3 \\ 
Dra      & dSph     & 86.37& $ 34.72$ & $ 79\pm~4$ & 4&0.74 &0.44 &$ -8.6$&$25.3\pm0.5$&$-2.1\pm0.4$ &34 \\
Sex      & dSph     &243.50& $ 42.27$ & $ 86\pm~6$ & 6&0.83 &0.52 &$ -9.5$&$26.2\pm0.5$&$-1.7\pm0.2$ &35 \\
Scl      & dSph     &287.54& $-83.16$ & $ 88\pm~4$ & 5&0.75 &0.45 &$ -9.8$&$23.7\pm0.4$&$-1.8\pm0.2$ &35 \\
Car      & dSph     &260.11& $-22.22$ & $ 94\pm~5$ & 7&0.82 &0.52 &$ -9.4$&$25.5\pm0.4$&$-2.0\pm0.2$ &36 \\
For      & dSph     &237.29& $-65.65$ & $138\pm~8$ & 8&0.76 &0.46 &$-13.1$&$23.4\pm0.3$&$-1.3\pm0.4$ &37 \\
Leo II   & dSph     &220.17& $ 67.23$ & $205\pm 12$& 9&0.87 &0.57 &$-10.1$&$24.0\pm0.3$&$-1.9\pm0.2$ &35 \\
Leo I    & dSph     &225.98& $ 49.11$ & $270\pm 10$&10&0.92 &0.63 &$-11.9$&$22.4\pm0.3$&$-1.5\pm0.3$ &35 \\
Phe      & dIrr/dSph&272.49& $-68.82$ & $405\pm 15$&11&0.85 &0.60 &$ -9.8$&   ---      &$-1.8\pm0.4$ &11 \\
NGC 6822 & IrIV-V   & 25.34& $-18.39$ & $500\pm 20$&12&0.91 &0.68 &$-16.0$&$21.4\pm0.2$&$-1.2\pm0.4$ &35 \\
\hline \\[-5pt]
M31      & SbI-II   &121.18& $-21.57$ & $770\pm 40$&13&0.00 &0.31 &$-21.2$&$10.8\pm0.4$& ---      &  \\
M32      & dE2,N    &121.15& $-21.98$ & $770\pm 40$&14&0.00 &0.31 &$-16.5$&$11.5\pm0.5$&$-1.1\pm0.6$ &35 \\
NGC 205  & dE5p,N   &120.72& $-21.14$ & $830\pm 35$&15&0.06 &0.37 &$-16.4$&$20.4\pm0.4$&$-0.5\pm0.5$ &38 \\
And I    & dSph     &121.69& $-24.85$ & $790\pm 30$&16&0.05 &0.33 &$-11.8$&$24.9\pm0.1$&$-1.5\pm0.2$ &39 \\
And III  & dSph     &119.31& $-26.25$ & $760\pm 70$&17&0.07 &0.30 &$-10.2$&$25.3\pm0.1$&$-1.5\pm0.2$ &39 \\
NGC 147  & dE5      &119.82& $-14.25$ & $755\pm 35$&18&0.10 &0.30 &$-15.1$&$21.6\pm0.2$&$-1.1\pm0.4$ &35 \\
And V    & dSph     &126.20& $-15.10$ & $810\pm 45$&19&0.12 &0.36 &$ -9.1$&$24.8\pm0.2$&$-1.9\pm0.1$ &39 \\
And II   & dSph     &128.87& $-29.17$ & $680\pm 25$&16&0.16 &0.24 &$-11.8$&$24.8\pm0.1$&$-1.5\pm0.3$ & 40 \\
NGC 185  & dE3p     &120.79& $-14.48$ & $620\pm 25$&20&0.17 &0.17 &$-15.6$&$20.1\pm0.4$&$-0.8\pm0.4$ &38 \\
M33      & ScII-III &133.61& $-31.33$ & $850\pm 40$&21&0.22 &0.42 &$-18.9$&$10.7\pm0.4$&   ---    & \\
Cas dSph & dSph     &109.46& $ -9.94$ & $760\pm 70$&22&0.22 &0.34 &$-12.0$&$23.5\pm0.1$&$-1.6\pm0.2$ &39 \\
IC 10    & IrIV:    &118.97& $ -3.34$ & $660\pm 65$&23&0.25 &0.26 &$-16.3$&$22.1\pm0.4$&$-1.3$: & 41 \\
And VI   & dSph     &106.01& $-36.30$ & $775\pm 35$&24&0.27 &0.38 &$-11.3$&$24.3\pm0.1$&$-1.9\pm0.2$ &39 \\
LGS 3    & dIrr/dSph&126.75& $-40.90$ & $810\pm 60$&25&0.28 &0.41 &$-10.5$&$24.7\pm0.2$&$-2.2\pm0.3$ & 42 \\
Peg      & IrV      & 94.77& $-43.55$ & $760\pm100$&26&0.41 &0.44 &$-12.3$&   ---      &$-1.3\pm0.3$ & 26 \\
IC 1613  & IrV      &129.82& $-60.54$ & $715\pm 35$&27&0.50 &0.47 &$-15.3$&$22.8\pm0.3$&$-1.4\pm0.3$ & 27 \\
\hline \\[-5pt]
Cet      & dSph     &101.50& $-72.90$ & $775\pm 50$&28&0.68 &0.62 &$-10.1$&$25.1\pm0.1$&$-1.9\pm0.2$ & 28 \\
Leo A    & IrV      &196.90& $ 52.40$ & $690\pm 60$&29&1.11 &0.88 &$-11.5$&   ---      &$-1.7\pm0.3$ & 29 \\
WLM      & IrIV-V   & 75.85& $-73.63$ & $945\pm 40$&30&0.84 &0.80 &$-14.4$&$20.4\pm0.1$&$-1.4\pm0.4$ & 30 \\
Tuc      & dSph     &322.91& $-47.37$ & $870\pm 60$&31&1.33 &1.11 &$ -9.6$&$25.1\pm0.1$&$-1.7\pm0.2$ & 31\\
DDO 210  & IrV      & 34.05& $-31.35$ & $950\pm 50$&32&1.08 &0.96 &$-10.9$&$23.0\pm0.3$&$-1.9\pm0.3$ & 32 \\
\hline \\
      \end{tabular}
  \end{center}
\end{table*}
  
\section{Deriving star formation histories}

\subsection{State of the art: Synthetic CMD techniques }
\label{sec:synth}

The derivation of star formation histories of resolved stellar populations
relies heavily on photometric imaging and color-magnitude diagrams (CMDs).  
The most detailed star formation histories are determined by modelling   
observed CMDs through the best-matched synthetic CMD (see \cite{Apar99} for
a review of synthetic CMD techniques).  The results depend on the quality 
and depth of the photometric input data, the treatment of observational
errors, the 
input evolutionary models, the sophistication of the modelling codes, etc. 
Free parameters include the choice of an IMF, binary fraction,
and enrichment law.  Due to different approaches and differing parameter
choices results obtained by different groups may not be directly comparable.
Some of the problems in deriving reliable star formation histories 
both on the theoretical and observational side are outlined below.

\subsection{Shortcomings of evolutionary models}
\label{sec:short}

While very impressive progress
has been made, models are limited by how well we understand 
the details of stellar evolution (e.g., mass loss in different phases of
stellar evolution, the impact of rotation and mixing, life times, blue-to-red
supergiant ratios, the second-parameter problem, etc.).  The currently used 
models are ``unphysical'' in the sense that they were calculated for 
non-rotating, single stars.  When the new rotational models become available it
will still be difficult to disentangle effects of rotation and binarity in
observational data.
Problems also persist in transforming theoretical models
to the observational plane, which can lead to discrepancies especially 
for cool stars (see Fig.\ 5 in \cite{Grebel99}).   

\subsection{Lack of abundance information}
\label{sec:noabund}

Broadband data (usually obtained in only
two filters) are insufficient to break the age-metallicity degeneracy 
particularly when dealing with complex, mixed populations.  
Metallicity-sensitive photometric systems work fairly well in populations with 
little age spread and can efficiently provide abundances (though not
necessarily iron abundances) for a large number of stars.  Spectroscopy 
is considered the most reliable method for abundance determinations,
but it is disquieting to consider the wide variety of spectroscopic abundances
found by various authors for the extensively studied young SMC cluster 
\object{NGC\,330}  (see \cite{Gonz99}).  

For most LG galaxies spectroscopic abundance information has either not been
obtained or is available only for very few evolutionary phases.  
[Fe/H] estimates are typically from the infrared Ca\,{\sc ii} triplet of
old red giants, from integrated spectra of globular clusters, or 
from luminous supergiants.  Photometric abundances are usually based on 
the photometric colors and slope of the red giant branch (\cite{Armand90}).
Oxygen abundance measurements exist for planetary nebulae (PNe)
or H\,{\sc ii} regions in about
half of the LG galaxies (\cite{Mateo98}).  
The relation between nebular abundances from H\,{\sc ii} regions
and PNe and stellar [Fe/H] values is poorly understood 
(\cite{Rich95}).  Discrepancies remain even when comparing element
abundances of young objects such as massive stars and H\,{\sc ii} 
regions (e.g., \cite{Garn99}).

\subsection{Area coverage}
\label{sec:area}

Most LG galaxies show evidence for
spatial variations in their star formation histories.  The sampling of
small subsections of a galaxy may not be representative for the overall
star formation history or lack the number statistics
to detect less well-populated age ranges (see \cite{Grebel97}).  This 
affects in particular old populations.  In more distant dIrrs
old populations were detected through deep ground-based observations
of their less crowded halos (\cite{Minn96};  
\cite{Minn99}).  Deep {\em HST} data of the center of Leo\,I led to the
suggestion of a lacking old population in this dSph (\cite{Gall99}), 
while wide-area imaging revealed a well-populated old horizontal branch
(\cite{Held00}).

\begin{figure}[ht]
  \begin{center}
    \epsfig{file=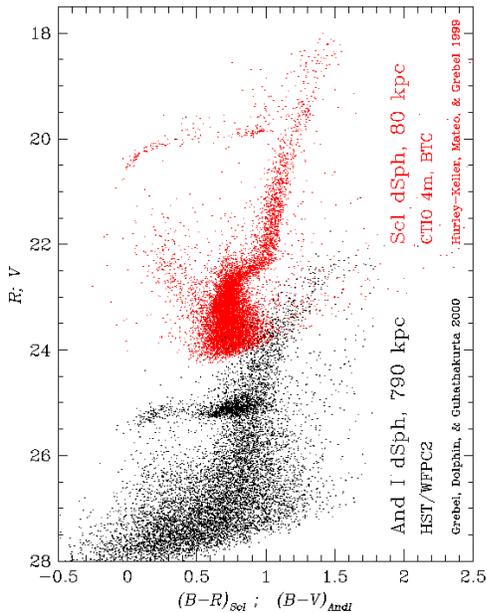, width=7cm}
  \end{center}
\caption{With increasing distance it becomes difficult to reach the 
main-sequence turnoff of the older populations, which is one of the
factors that contribute to decreased
resolution for older ages.  In dIrrs superimposed younger main
sequence stars aggravate the problem and may not only make the oldest
main-sequence turnoff and subgiant branch hard to measure but also 
merge with the old blue horizontal branch.
\label{fig2}}
\end{figure}

\subsection{Age resolution}
\label{sec:age_resol}

While we can determine the ages for young
populations traced by massive stars to a few $10^6$ to $10^7$ years,
the age resolution decreases roughly logarithmically for older populations 
(see Fig.\ 4 in \cite{Grebel99}) within a given galaxy.  
Furthermore, the attainable age resolution decreases with distance from
the Milky Way (Fig.\ \ref{fig2}).

\section{Star formation in the three spirals}
\label{sec:spiral}

A detailed description of the star formation histories of the three 
spirals in the LG
is beyond the scope of this short paper.  Instead I will attempt
a brief summary and refer to the recent reviews by \cite*{Maj99} and
van den Bergh (1999b; 2000a) for more information.  

The star formation histories of spiral galaxies vary with position
within the galaxy depending on which galaxy component is considered -- 
halo, bulge, or disk, and further morphological or kinematic subdivisions. 

\subsection{The Milky Way}
\label{sec:MW}

The Milky Way is a S(B)bc galaxy with continuing star formation
in the Galactic center (\cite{Figer99}) and in its spiral arms.
The oldest components of the Milky Way are the old halo (both field
stars and old globulars) and the thick disk.  The formation of the old
halo may be best described by \cite*{SZ78}-like accretion.
Most of the merger events are believed to be ancient
(\cite{Una96}).  Efforts are on the way to trace their remnants by
surveying the Galactic halo for kinematic substructure (e.g., \cite{Morris00}).

The earliest significant star formation in the thick disk 
may date back 13 Gyr ago.  The thick disk has a mean metallicity of
$-0.6$ dex (\cite{Maj99}) and a dynamically hot, extended structure.
The thin disk experienced multiple bursty star formation episodes starting 
$\sim 9$ Gyr ago.  \cite*{Roch00}
conclude that the global disk star formation rate in the Milky Way fluctuates
on time scales of less than 0.2 -- 1 Gyr with amplitudes exceeding a factor
2 -- 3.  They suggest that such intermittent star formation may be typical
for disk galaxies though for the Milky Way interactions with
the Magellanic Clouds may have played a role as well. 
While there are indications of a radial abundance gradient
in the thin disk the metallicity is a stronger function of galactocentric
radius than of age and exhibits large scatter at any position and age 
(\cite{Edvard93}).
Mean metallicities range from $-0.5 < [{\rm Fe/H}] < 0.1$ dex with a pronounced
metal-poor tail.  One of the scenarios for thin-disk evolution is inside-out
formation (\cite{Edvard93}), i.e., the inner parts form more rapidly and
experience stronger enrichment.   

The Galactic bulge as well as other bulges seem to have undergone early
and very rapid enrichment.  The mean metallicity of the bulge is $-0.25$ dex
with a spread that can easily exceed 1 dex (\cite{Minn95}).  Also, a radial
abundance gradient is observed.  Both the abundance spread and wide range 
in ages may be in part caused by contributions from overlapping 
Milky Way halo and disk components.  Bulge formation in 
a rapid monolithic collapse similar to the scenario
suggested by \cite*{Eggen62} appears likely.

\subsection{M31}
\label{sec:M31}
 
M31, an Sb I--II spiral, usually believed to be $\sim 30$\% more massive than
the Milky Way (\cite{Wilk99}), may in fact be only as massive or even less
massive than the Milky Way (\cite{Evans00}).  It
shows on average higher metallicities than the Milky Way and a considerable 
abundance spread.  The outer halo has a mean metallicity of $-1.2$ dex,
while for the inner halo $\rm\langle[Fe/H]\rangle = -1.0$ dex 
(\cite{vdB00} and references therein).  M31 appears to have undergone
rapid enrichment as a whole.  The bulge emits $\sim 30\%$ of the visible
light of M31 and contains two nuclei, which may be both part of the same
eccentric disk of stars orbiting a central black hole (\cite{Korm99}).
M31 has a warped stellar and gas
disk, possibly caused 
by tidal interactions with massive nearby companions such as M32.
The horizontal branch morphology of some of the $\sim 600$ globular 
clusters in M31 suggests that they have similarly ages as the oldest Galactic
globulars
(e.g., \cite{Ajhar96}).  M31 currently only shows low star formation activity.
The dominant present-day star formation is occurring in the outer disk.
The UV line strengths of massive OB stars in M31 indicate a metallicity 
comparable to that of the young Milky Way population (\cite{Bia96}).

\subsection{M33}
\label{sec:M33}

M33 is an Sc II--III spiral with a significant radial abundance gradient.
It shows a nucleus, a pronounced disk and halo, but no obvious bulge.
The mean metallicity of M32's halo ($-1.6$ dex) is lower than M31's and
shows a spread.  
Stellar and H\,{\sc i} disk are tilted with respect to each other, but M33 
doesn't have any known companions that might be responsible.  About 
54 globular clusters are currently known in M33 (\cite{Moch98}).
The lack of a blue horizontal branch (BHB) in M33 globulars may suggest that
the bulk of M33's star formation began several Gyr later than in the Milky Way
(\cite{Sara98}).  In the past $10^7$ to $10^8$ years cluster formation 
increased (\cite{Chan99}) and appears to be correlated with gas inflow into
the center of M33.  The UV line strengths of massive OB stars in M33
resemble the young LMC population in metallicity (\cite{Bia96}).  

\section{Stellar luminosity functions}
\label{sec:LFs}
 
The behavior of the high-mass portion of the stellar luminosity function (LF)
and initial mass function (IMF) is directly measurable only in well-resolved
young populations.  The careful spectroscopic and photometric studies by
Massey et al.\ (1995a,b) of OB associations in the Milky
Way, LMC, and SMC showed that the upper IMFs vary little and are essentially 
consistent with a Salpeter IMF within the errors despite the differences
in metallicity.  This was confirmed
for a range of masses down to 2.8 -- 0.7 M$_{\odot}$ 
for young clusters in the LMC (\cite{Mass98}; \cite{Elson99}; 
\cite{GrChu00}).  The only known exceptions in single-age populations
are the Galactic center clusters
and the central starburst cluster in the Galactic giant H\,{\sc ii} region
NGC\,3603, which seem to have significantly flatter IMFs (\cite{Figer99};
\cite{Eisen98}).  
 
The LFs of older, low-mass field populations in the
Galactic bulge (\cite{Holtz98}), LMC and SMC (\cite{Holtz99b}),
Draco (\cite{Grill98}), and Ursa Minor (\cite{Felt99}) are 
in excellent agreement with the solar neighborhood IMF or
that of globular clusters unaffected by significant mass segregation.
 
Apart from providing support for the universality of the IMF the
findings for dSphs show that low-mass stars down to masses of 0.45
M$_{\odot}$ are not responsible for the inferred high mass-to-light
ratios of these galaxies.

\section{The oldest populations in the Local Group}

Old populations ($> 10$ Gyr) have been detected in all LG galaxies studied
in sufficient detail.  Tracers of the oldest populations include
old main-sequence turnoffs and subgiant branches, horizontal branches, 
and RR Lyrae in globular clusters and field populations.

Deep imaging largely based on {\em HST} data has established that a number
of LG galaxies share a common epoch of ancient star formation.  Main-sequence
photometry reveals that the oldest globular clusters in the Milky Way halo
and bulge,
in the LMC (\cite{Olsen98}), in Sagittarius (\cite{Monte98}), and in Fornax
(\cite{Buon98}) are coeval within the measurement accuracy. Similarly,
deep main-sequence photometry of the oldest field populations in 
Sagittarius (\cite{Layd00}), LMC (\cite{Holtz99}), Draco, Ursa Minor
(\cite{Migh99}), Fornax, Sculptor (\cite{Monk99}), Carina (\cite{Migh97}),
and \object{Leo\,II} (\cite{Migh96}) indicates that they have the same
relative age as the oldest Galactic globular clusters.  

Though main-sequence photometry is lacking, the CMD of  
the globular cluster in the \object{WLM} dIrr ({\cite{Hodge99}),
the BHBs of globulars in M31 (\cite{Ajhar96}) and in 
NGC\,147 (\cite{Han97}), and spectroscopy of one of NGC 6822's 
globulars (\cite{Cohen98}) are interpreted as indicative of ages 
similar to those of the old Galactic globular clusters.  Comparatively old
ages may be implied by the 
existence of a BHB in the field populations of Sextans
(\cite{Mateo91}), Leo\,I (\cite{Held00}), Phoenix (\cite{Smith00}),
IC\,1613 (\cite{Cole99}, Cetus (\cite{Tols00}),
And\,I (\cite{DaCosta96}), And\,II (\cite{DaCosta00}), NGC\,185 
(\cite{Geis99}), NGC\,147 (\cite{Han97}), and Tucana (\cite{Lave96}).  
However, while a BHB indicates an old population it should be emphasized 
that due to second-parameter effects other than age a BHB does not
not necessarily imply an ancient, old halo globular-cluster-like population.

Conversely (and with the same caveat), the absence of a BHB in M33 globulars 
(\cite{Sara98}) and the
apparent lack of a pronounced BHB in the field populations of the SMC, WLM 
(\cite{Dolp00}), Leo\,A (\cite{Tols98}), and DDO\,210 (\cite{Tols00})
may indicate that the bulk of the old population in these galaxies
formed with a delay of a few Gyr.

\section{Spatial variations and tidal effects}
\label{sec:spatvar}

\subsection{Spatial variations in star formation history}
\label{sec:spatvar_SFH}

Spatial variations in star formation history are a common occurrence 
in galaxies (see also Section \ref{sec:area}) and provide clues
about the distribution of the star-forming material as a function of
time, the dynamical history of galaxies as well as of the gas retention 
or gas loss history.  The extent of spatial variations seems to be
correlated both with the overall mass of the galaxy and its environment.

A global property of LG galaxies 
is that the oldest populations are the most extended ones.    
This may indicate that galaxies have contracted with time (\cite{vdB99b})
or may be a consequence of dispersion or substantial mass loss.  
High-mass dwarf galaxies
such as dIrrs typically show multiple distinct zones of concurrent star
formation.  Some of these regions can be long-lived, exist for several 
100 Myr, and may migrate (e.g., Sextans A: \cite{Dohm97}; LMC: \cite{GrBr99}).  
In contrast, low-mass satellites (especially dSphs) tend to have centrally 
concentrated intermediate-age or young populations and may show radial
age gradients.  Star-forming
material appears to have been retained longer in the central regions.
An example of a dSph with a very pronounced age and metallicity gradient 
is Fornax (\cite{Stet98}; \cite{GrSt99}).  In Fornax different populations
have slightly different centroids, and the youngest population is 
asymmetrically distributed.

In some cases spatial variations
in star formation history are traced only by horizontal-branch variations
in the sense that BHB stars are more spatially extended than red horizontal
branch stars (examples include Sculptor (\cite{Hurl99}) and And\,I
(\cite{DaCosta96})).  
A few low-mass dwarf galaxies show no apparent 
indication for spatial population variations including Ursa Minor, Carina 
(\cite{Smec96}), and And\,II ({\cite{DaCosta00}).  

\subsection{Spatial variations in stellar density}
\label{sec:spatvar_dens}

While lacking variations in its stellar populations, Ursa Minor shows
statistically significant stellar density variations (\cite{Kleyna98})
instead of a smooth surface density profile.  
In the more luminous dSph Fornax four out of the five  globular clusters 
are found at distances larger than the galaxy's core radius although
their orbital decay scales through dynamical friction
are only a few Gyr (\cite{Oh00}), a fraction of the globular clusters'
ages.  
Significant mass loss through Galactic tidal perturbation and the resulting 
decrease in the satellite galaxy's gravitational potential 
may have increased the clusters' orbital semimajor axes and efficiently
counteracted the spiralling-in through dynamical friction (\cite{Oh00}).

The currently available radial velocity measurements for individual stars
in dSphs do not yet allow one to distinguish between dynamical models 
where mass follows light, extended dark halo models, and disrupted remnant
models without dark matter 
(\cite{Kleyna99}).  Interestingly, the detection of a possible
very extended population of extratidal stars around the dSph Carina
might imply that this galaxy has now been reduced to a mere 1\% of its
initial mass (\cite{Maj00}), the rest of which has already been accreted
by the Milky Way's halo.  Apart from favoring disrupted remnant models 
this claim has far-reaching consequences for the determination of
star formation histories of nearby low-mass satellites:
If they are indeed significantly tidally disrupted 
then their present-day stellar content cannot easily be used to
derive evolutionary histories over a Hubble time. 
The observed metallicity spread in the nearby dSph Draco and other dSphs
provides further evidence
for significant past mass loss since this dwarf was once massive enough
to retain its metals for subsequent star formation.  According to 
\cite*{Mac99} almost no metals are retained by galaxies with 
$< 10^9$ M$_{\odot}$.  

Substantial tidal disruption by the Milky Way 
is evidenced by the Magellanic Clouds and
Magellanic Stream as well as by the Sagittarius dSph.  
Tracer features include gaseous and stellar tidal tails, kinematics, and
structural parameters (e.g., \cite{Hatz93}, \cite{Putman98}, \cite{Maj99b}, 
\cite{Wein00}, \cite{Ibata00}).  The dSphs Draco and Ursa Minor as well
as various globular clusters may be part of the extended Magellanic Stream
(\cite{Kunkel79}).  Peaks in the star cluster formation rate could be 
related to close encounters between the Magellanic Clouds and the Milky
Way (\cite{Gir95}, \cite{GrZa99}).  The stellar metallicity patterns in the
intercloud region impose time constraints on the tidal disruption 
(\cite{Roll99}). 

\subsection{Variations in the gas content of dwarfs}
\label{sec:spatvar_gas}

The gas in dwarf galaxies shows density and spatial variations as well.  

The H\,{\sc i} in dIrrs tends to be even more extended than the oldest 
stellar populations and shows a clumpy distribution.
Central H\,{\sc i} holes or off-centered gas
concentrations are observed in a number of cases and may be driven by
star formation events (Young \& Lo 1996; 1997a,b). %; \cite{Stan99}).    
Ongoing gas accretion has been found in at least one dIrr (\object{IC\,10}: 
\cite{Wil98}).  Low amounts of gas ($\sim 10^5 M_{\odot}$) were detected
in dEs (\cite{Sage98}).  The gas kinematics of \object{NGC\,185} indicate
an internal origin (i.e., consistent with stellar gas loss), while the 
significant angular momentum of the H\,{\sc i} in \object{NGC\,205} 
as compared to the non-rotating stellar component may gas imply accretion.
Single-dish observations of dSphs did not detect gas (e.g., \cite{Knapp78}),
and subsequent VLA synthesis studies covering much of the optical extent
of some of the dSphs did not reveal H\,{\sc i} in emission or in 
absorption down to column densities of $10^{17}$ -- $10^{18}$ cm$^{-2}$
({\cite{Young99}; 2000).   

The apparent absence of gas in dSphs is 
an unsolved puzzle.  Simulations by \cite*{Mac99} indicate that gas loss
through starbursts becomes efficient only at masses $< 10^{-6}$,
an order of magnitude less than the typical mass of a dSph.  The observed
upper H\,{\sc i} limits are lower than even what is expected from evolutionary
mass loss from red giants in dSphs.  The lack of gas is  
particularly puzzling in dSphs and dEs with pronounced intermediate-age
populations (Carina: 3 Gyr, \cite{Hurl98}; Leo\,I: 2 Gyr,
\cite{Gall99}; NGC\,147, \cite{Han97}), or very recent 
star formation (Fornax: $\sim 200$ Myr, \cite{GrSt99}).  
However, \cite*{Car98} and \cite*{Car99} found extended gas lobes
around the Sculptor dSph, whose radial velocities are similar to the
stellar radial velocity of this galaxy.  A recent re-investigation of
the Leiden-Dwingeloo survey led to the detection of similar gas 
concentrations with matching radial velocities outside the optical boundaries
of several other dSphs and to non-detections for
others (\cite{Blitz00}).  These authors suggest that tidal effects are the
most likely agent for the displacement of the gas. 

\section{Star Formation in Local Group Dwarfs}

As illustrated in earlier reviews (\cite{Grebel97}, 1998, \cite{Mateo98},
\cite{Grebel99})
dwarf galaxies vary widely in their star formation histories, age 
distribution, and enrichment history.  No two dwarfs are alike even
within the same morphological type.
In spite of this diversity, galaxy mass and proximity to a massive spiral
appear to play a defining role in dwarf galaxy evolution.    

\subsection{Modes of star formation}

The following modes of star formation are represented among LG dwarfs: 
\begin{itemize}
\item Continuous star formation with a constant or varying star formation
rate over a Hubble time and gradual enrichment
(see also \cite{Hunter97}).
Examples for this mode include irregulars and dIrrs such as the
Magellanic Clouds, which are massive enough to have a sufficiently large
gas supply and to hold on to gas and metals.  
\item Continuous star formation with decreasing star
formation rate that ceases eventually.   A good example is the Fornax dSph
(e.g., \cite{GrSt99}).  This mode may be dominant in low-mass
dIrrs and transition-type galaxies as well.  External effects such as
tidal or ram-pressure stripping may have contributed to the gradual loss
of star-forming material in these galaxies.  Extreme cases at the low-mass
end are two of the closest and least massive Milky Way dSph companions, Draco
and Ursa Minor, which are dominated by ancient populations (see also
Section \ref{sec:LFs}).  However, the
abundance spread found in Draco (\cite{Shet98})
and the presence of a few Carbon stars indicates that the early star
formation episodes must have been fairly extended. 
\item Distinct star formation episodes separated by Gyr-long periods of 
quiescence.  So far only one example of this mode is known, the Carina dSph 
(e.g., \cite{Hurl98}).  It is unclear what caused the gaps and the 
subsequent onset of star formation.  One possible explanation is the
episodic scenario proposed by \cite{Lin99}:  star formation in a dwarf
heats and disperses the gas, which remains bound and eventually
cools and contracts during the apogalactic passages of the dwarf around
the massive parent.  According to their simulations this leads to star
formation separated by gaps of a few Gyr. --- 
The distinct episodes of star formation in Carina 
seem to have proceeded without enrichment,
a possible indication of star formation through accretion or else metal
loss as described by \cite*{Mac99}. 
\end{itemize}

Our present lack of knowledge of dwarf galaxy orbits makes it impossible
to evaluate the impact of past close encounters and interactions on dwarf
galaxy star formation histories.

\subsection{Evolution from dIrrs to dSphs}

\begin{figure}[htb]
  \begin{center}
    \epsfig{file=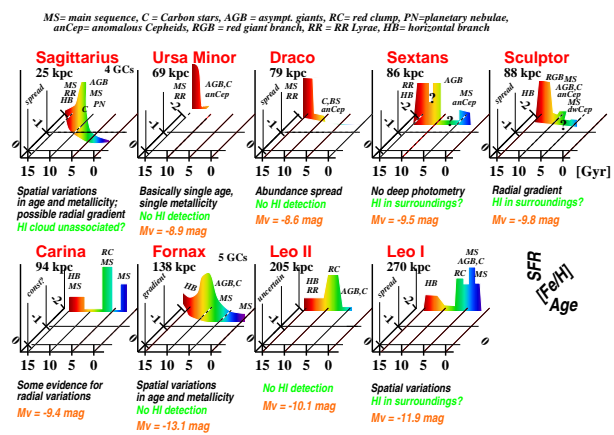, width=8cm}
  \end{center}
\caption{Star formation histories of Milky Way dwarf spheroidal 
         companions.  Each population box gives a schematic representation
         of star formation rate (SFR) as a function of age and metallicity.
         The fraction of intermediate-age populations tends to increase with 
         increasing Galactocentric distance and dwarf galaxy mass.  On the
         other hand, the distant, isolated dSph Tucana appears to have 
         predominantly old populations.
\label{fig3}}
\end{figure}

For the Milky Way dSph companions there is a general tendency for increasing
intermediate-age fractions with increasing Galactocentric distance,
which appears to support ram pressure and tidal stripping (\cite{Ein74};
\cite{Lin83}, \cite{vdB94}) as major forces.

All morphological types of LG dwarf galaxies tend to  follow global 
relations between absolute
magnitude, mean metallicity, and central surface brightness.  The more
luminous a galaxy the higher its metallicity (e.g., \cite{GrGu99}).  
These trends
indicate the importance of dwarf galaxy mass on its evolution and 
ability to retain metals. 

The presence of intermediate-age or even young populations in some of the
more distant dSphs, the possible detection of associated gas in the 
surroundings of some of the dSphs, indications of substantial mass loss
(Section \ref{sec:spatvar_dens}), morphological segregation, common
trends in relations between their integrated properties, and the
apparent correlation between star formation histories and Galactocentric
distance all seem to support the idea that low-mass dIrrs will eventually
evolve into dSphs, which may be fostered by their environment.  In fact
the distinction between dSphs and dIrrs may be more semantical than 
physical.  

Similarly detailed information on the M31 dSphs, which span the same range
of distances from M31 as the Milky Way dSph companions from the Milky Way,
would help to evaluate this scenario (\cite{GrGu99}). 

On the other hand, this cannot account for the existence of isolated,
predominantly old dSphs such as Tucana.  Tucana's large distance from
the Milky Way and M31 (see Table 2) lets it appear unlikely that this
dSph suffered major interactions in the past.

\section{Summary}

Star formation histories in the LG differ from galaxy to galaxy and
vary widely in length and times of their star formation episodes and
enrichment history.  In spite of their diversity galaxy mass and 
environment are important factors in determining the star formation
histories.  Old populations have been found in all LG galaxies studied
in detail so far, though in some LG galaxies measurable star formation
appears to have begun a few Gyr later than in others.  Spatial
variations in ages and abundances are observed in most LG galaxies.
Low-mass dwarfs tend to exhibit radial population gradients with
extended old populations and centrally concentrated young stars.  Global
modes of star formation range from continuous (variable star formation
rate or gradual decrease) to episodic star formation separated by long
periods of quiescence.  Again, this appears to depend partially on 
galaxy mass and environment.  Integrated properties such as mean 
metallicity, central surface brightness, and absolute magnitude follow
common trends for all LG dwarfs.  
The idea that low-mass dIrrs may eventually evolve into
dSphs is consistent with the range of observed properties of transition-type
dwarfs and dSphs.  Major unsolved questions include the fate of the gas in
dSphs, whether dSphs have been substantially affected by mass loss, 
unknown satellite orbits, and the problem of the ``missing'' LG satellites.

\begin{acknowledgements}

This work was supported by NASA through grant HF-01108.01-98A from the
Space Telescope Science Institute, which is operated by the Association of
Universities for Research in Astronomy, Inc., under NASA contract NAS5-26555.
I gratefully acknowledge support from the organizers through an ESA grant.

It is a pleasure to thank Robert Braun, Kem Cook, Sandy Faber, 
Jay Gallagher, Raja Guhathakurta, Paul Hodge, and Igor Karachentsev 
for valuable discussions. 

\end{acknowledgements}


\begin{thebibliography}{}

\bibitem[\protect\astroncite{Ajhar et~al.}{1996}]{Ajhar96}
Ajhar, E.A., Grillmair, C.J., Lauer, T.R., et al. 1996, AJ, 111, 1110

\bibitem[\protect\astroncite{Aparicio}{1999}]{Apar99}
Aparicio, A., 1999, in The Stellar Content of the Local Group,
IAU Symposium 193, eds. P. Whitelock \& R. Cannon, Astronomical
Society of the Pacific

\bibitem[\protect\astroncite{Armandroff \& Da Costa}{1990}]{Armand90}
Armandroff, T.E., Da Costa, G.S., 1990, AJ, 100, 162

\bibitem[\protect\astroncite{Armandroff et~al.}{1993}]{Armand93}
Armandroff, T.E., Da Costa, G.S., Caldwell, N., Seitzer, P., 1993, AJ, 106, 986

\bibitem[\protect\astroncite{Armandroff et~al.}{1998}]{Armand98}
Armandroff, T.E., Davies, J.E., \& Jacoby, G.H., 1998, AJ, 116, 2287

\bibitem[\protect\astroncite{Armandroff et~al.}{1999}]{Armand99}
Armandroff, T.E., Davies, J.E., \& Jacoby, G.H., 1999, AJ, 118, 1220

\bibitem[\protect\astroncite{Bellazzini et~al.}{1999}]{Bella99}
Bellazzini, M., Ferraro, F.R., Buonanno, R., 1999, MNRAS, 307, 619

\bibitem[\protect\astroncite{Bianchi et~al.}{1996}]{Bia96}
Bianchi, L., Hutchings, J.B., Massey, P., 1996, AJ, 111, 2303

\bibitem[\protect\astroncite{Blitz \& Robishaw}{2000}]{Blitz00}
Blitz, L., Robishaw, T., 2000, ApJ, submitted (astro-ph/0001142)

\bibitem[\protect\astroncite{Braun \& Burton}{1999}]{Braun99}
Braun, R., Burton, W.B., 1999, A\&A, 341, 437

\bibitem[\protect\astroncite{Braun \& Burton}{2000}]{Braun00}
Braun, R., Burton, W.B., 2000, A\&A, 354, 853

\bibitem[\protect\astroncite{Buonanno et~al.}{1998}]{Buon98}
Buonanno, R., Corsi, C.E., Zinn, R., et al., 1998, ApJ, 501, L33

\bibitem[\protect\astroncite{Caldwell}{1999}]{Cald99}
Caldwell, N., 1999, AJ, 118, 1230

\bibitem[\protect\astroncite{Carignan}{1999}]{Car99}
Carignan, C., 1999, PASA, 16, 18

\bibitem[\protect\astroncite{Carignan}{1998}]{Car98}
Carignan, C., Beaulieu, S., C\^ot\'e, S., et al., 1998, AJ, 1690

\bibitem[\protect\astroncite{Chandar et~al.}{1999}]{Chan99}
Chandar, R., Bianchi, L., Ford, H.C., 1999, ApJ, 517, 668

\bibitem[\protect\astroncite{Cohen \& Blakeslee}{1998}]{Cohen98}
Cohen, J.G., Blakeslee, J.P., 1998, AJ, 115, 2356

\bibitem[\protect\astroncite{Cole et~al.}{1999}]{Cole99}
Cole, A.A., et al., 1999, AJ, 118, 1657

\bibitem[\protect\astroncite{Cook et~al.}{1999}]{Cook99}
Cook, K.H., Mario, M., Olszewski, E.W.,
Vogt, S.S., Stubbs, C., Diercks, A., 1999, PASP, 111, 306

\bibitem[\protect\astroncite{C\^ot\'e et~al.}{1999}]{Cote99}
C\^ot\'e, P., Oke, J.B., Cohen, J.G., 1999, AJ, 118, 1645
 
\bibitem[\protect\astroncite{Courteau \& van den Bergh}{1999}]{Court99}
Courteau, S., van den Bergh, S., 1999, AJ, 118, 337

\bibitem[\protect\astroncite{Da Costa et~al.}{1996}]{DaCosta96}
Da Costa, G.S., Armandroff, T.E., Caldwell, N., Seitzer, P., 1996,
AJ, 112, 2576

\bibitem[\protect\astroncite{Da Costa et~al.}{2000}]{DaCosta00}
Da Costa, G.S., Armandroff, T.E., Caldwell, N., Seitzer, P., 2000,
AJ, 119, 705

\bibitem[\protect\astroncite{Djorgovski et~al.}{1999}]{Djor99}
Djorgovski, S.G., Gal, R.R., Odewahn, S.C., et al. 1999, 
in Wide Field Surveys in Cosmology, ed. S. Colombi et al., Editions
Fronti\`eres

\bibitem[\protect\astroncite{Dohm-Palmer et~al.}{1997}]{Dohm97}
Dohm-Palmer, R.C., Skillman, E.D., Saha, A., et al., 1997, AJ, 114, 2527

\bibitem[\protect\astroncite{Dolphin}{2000}]{Dolp00}
Dolphin, A.E., 2000, ApJ, 531, 804

\bibitem[\protect\astroncite{Dressler}{1980}]{Dressler80}
Dressler, A. 1980, ApJ, 236, 351

\bibitem[\protect\astroncite{Dressler et~al.}{1997}]{Dressler97}
Dressler, A., Oemler, A., Warrick, J.C., et al. 1997, ApJ, 490, 577

\bibitem[\protect\astroncite{Feltzing et~al.}{1999}]{Felt99}
Feltzing, S., Gilmore, G., Wyse, R.F.G., 1999, ApJ, 516, L17

\bibitem[\protect\astroncite{Edvardsson et~al.}{1993}]{Edvard93}
Edvardsson, B., Andersen, J., Gustafsson, B., et al., 1993, A\&A, 273, 101

\bibitem[\protect\astroncite{Eggen et~al.}{1962}]{Eggen62}
Eggen, O.J., Lynden-Bell, D., Sandage, A.R., 1962, ApJ, 136, 748

\bibitem[\protect\astroncite{Einasto et~al.}{1974}]{Ein74}
Einasto, J., Saar, E., Kaasik, A., Chernin, A.D., 1974, Nature, 252, 111

\bibitem[\protect\astroncite{Eisenhauer et~al.}{1998}]{Eisen98}
Eisenhauer, F., Quirrenbach, A., Zinnecker, H., Genzel, R., 1998, ApJ, 498, 278

\bibitem[\protect\astroncite{Elson et~al.}{1999}]{Elson99}
Elson, R., Tanvir, N., Gilmore, G., et al., 1999, in
New Views of the Magellanic Clouds, IAU
Symposium 190, eds. Y.-H. Chu et al., Astronomical
Society of the Pacific

\bibitem[\protect\astroncite{Evans et~al.}{2000}]{Evans00}
Evans, N.W., Wilkinson, M.I., Guhathakurta, P., et al., 2000, ApJ Lett, subm.

\bibitem[\protect\astroncite{Gallagher \& Wyse}{1994}]{Gall94}
Gallagher, J.S., Wyse, R.F.G., 1994, PASP, 106, 1225

\bibitem[\protect\astroncite{Gallagher et~al.}{1998}]{Gall98}
Gallagher, J.S., Tolstoy, E., Dohm-Palmer, R.C., et al.,
1998, AJ, 115, 1869

\bibitem[\protect\astroncite{Gallart et~al.}{1996}]{Gall96}
Gallart, C., Aparicio, A., V\'{\i}lchez, J.M., 1996, AJ, 112, 1950

\bibitem[\protect\astroncite{Gallart et~al.}{1999}]{Gall99}
Gallart, C., Freedman, W.L., Aparicio, A., et al.,
1999, AJ, 118, 2245

\bibitem[\protect\astroncite{Garnett}{1999}]{Garn99}
Garnett, D.R., 1999, in New Views of the Magellanic Clouds, IAU 
Symposium 190, eds. Y.-H. Chu et al., Astronomical
Society of the Pacific

\bibitem[\protect\astroncite{Geisler et~al.}{1999}]{Geis99}
Geisler, D., Armandroff, T.E., Da Costa, G., et al.,
1999, in The Stellar Content of the Local Group,
IAU Symposium 193, eds. P. Whitelock \& R. Cannon, Astronomical
Society of the Pacific

\bibitem[\protect\astroncite{Girardi et~al.}{1995}]{Gir95}
Girardi, L., Chiosi, C., Bertelli, G., Bressan, A., 1995, A\&A, 298, 87

\bibitem[\protect\astroncite{Gonz\'alez \& Wallerstein}{1999}]{Gonz99}
Gonz\'alez, G., Wallerstein, G., 1999, AJ, 117, 2286

\bibitem[\protect\astroncite{Grebel}{1997}]{Grebel97}
Grebel, E.K., 1997, Rev.\ in Mod.\ Astron., 10, 29

\bibitem[\protect\astroncite{Grebel}{1999}]{Grebel99}
Grebel, E.K., 1999, in The Stellar Content of the Local Group, 
IAU Symposium 193, eds. P. Whitelock \& R. Cannon, Astronomical
Society of the Pacific

\bibitem[\protect\astroncite{Grebel}{2000}]{Grebel00}
Grebel, E.K., 2000, in Massive Stellar Clusters, Workshop of the
Observatoire de Strasbourg, eds.\ A. Lan\c{c}on \& C. Boily, 
Astronomical Society of the Pacific

\bibitem[\protect\astroncite{Grebel \& Brandner}{1999}]{GrBr99}
Grebel, E.K., Brandner, W., 1999, in
New Views of the Magellanic Clouds, IAU
Symposium 190, eds. Y.-H. Chu et al., Astronomical
Society of the Pacific

\bibitem[\protect\astroncite{Grebel \& Guhathakurta}{1999}]{GrGu99}
Grebel, E.K., Guhathakurta, P., 1999, ApJ, 511, L101

\bibitem[\protect\astroncite{Grebel \& Stetson}{1999}]{GrSt99}
Grebel, E.K., Stetson, P.B., 1999, in The Stellar Content of the Local Group,
IAU Symposium 193, eds. P. Whitelock \& R. Cannon, Astronomical
Society of the Pacific

\bibitem[\protect\astroncite{Grebel et~al.}{1999}]{GrZa99}
Grebel, E.K., Zaritsky, D., Harris, J., Thompson, I., 1999, in
New Views of the Magellanic Clouds, IAU
Symposium 190, eds. Y.-H. Chu et al., Astronomical
Society of the Pacific

\bibitem[\protect\astroncite{Grebel \& Chu}{2000}]{GrChu00}
Grebel, E.K., Chu, Y.-H., 2000, AJ, 119, 787

\bibitem[\protect\astroncite{Grillmair et~al.}{1996}]{Grill96}
Grillmair, C.J., Lauer, T.R., Worthey, G., et al., 1996, AJ, 112, 1975

\bibitem[\protect\astroncite{Grillmair et~al.}{1998}]{Grill98}
Grillmair, C.J., Mould, J.R., Holtzman, J.A., et al. 1998, AJ, 115, 144

\bibitem[\protect\astroncite{Guhathakurta et~al.}{2000}]{Guha00}
Guhathakurta, P., Grebel, E.K., Pittroff, L.C., et al., 2000, AJ, in prep.

\bibitem[\protect\astroncite{Gunn \& Weinberg}{1995}]{Gunn95}
Gunn, J.E., Weinberg, D.H., 1995, in Wide Field Spectroscopy and the Distant
Universe, ed. S. Maddox \& A. Arag\'on-Salamanca, World Scientific 

\bibitem[\protect\astroncite{Ferguson \& Binggeli}{1994}]{Ferg94}
Ferguson, H.C., Binggeli, B., 1994, A\&ARv, 6, 67

\bibitem[\protect\astroncite{Figer et~al.}{1999}]{Figer99}
Figer, D.F., Kim, S.S., Morris, M., et al., 1999, ApJ, 525, 750

\bibitem[\protect\astroncite{Freedman \& Madore}{1990}]{Freed90}
Freedman, W.L., Madore, B.F., 1990, ApJ, 365, 186

\bibitem[\protect\astroncite{Freedman et~al.}{1991}]{Freed91}
Freedman, W.L., Wilson, C.D., Madore, B.F., 1991, ApJ, 372, 455

\bibitem[\protect\astroncite{Han et~al.}{1997}]{Han97}
Han, M., Hoessel, J.G., Gallagher, J.S., et al., 1997, AJ, 113, 1001

\bibitem[\protect\astroncite{Hatzidimitriou et~al.}{1993}]{Hatz93}
Hatzidimitriou, D., Cannon, R.D., Hawkins, M.R.S., 1993, MNRAS, 261, 873

\bibitem[\protect\astroncite{Held et~al.}{1999}]{Held99}
Held, E.V., Saviane, I., Momany, Y., 1999, A\&A, 345, 747

\bibitem[\protect\astroncite{Held et~al.}{2000}]{Held00}
Held, E.V., Saviane, I., Momany, Y., Carraro, G., 2000, ApJ, 530, L85

\bibitem[\protect\astroncite{Hodge et~al.}{1999}]{Hodge99}
Hodge, P.W., Dolphin, A.E., Smith, T.R., Mateo, M., 1999, ApJ, 521, 577

\bibitem[\protect\astroncite{Holtzman et~al.}{1998}]{Holtz98}
Holtzman, J.A., Watson, A.M., Baum, W.A., et al., 1998, AJ, 115, 1946

\bibitem[\protect\astroncite{Holtzman et~al.}{1999a}]{Holtz99}
Holtzman, J.A., Gallagher, J.S., Cole, A.A., et al., 1999, AJ, 118, 2262

\bibitem[\protect\astroncite{Holtzman et~al.}{1999b}]{Holtz99b}
Holtzman, J.A., Mould, J.R., Gallagher, J.S., 1999b, in
New Views of the Magellanic Clouds, IAU
Symposium 190, eds. Y.-H. Chu et al., Astronomical
Society of the Pacific

\bibitem[\protect\astroncite{Hunter}{1997}]{Hunter97}
Hunter, D.A., 1997, PASP, 109, 937

\bibitem[\protect\astroncite{Hurley-Keller et~al.}{1998}]{Hurl98}
Hurley-Keller, D., Mateo, M., Nemec, J., 1998, AJ, 115, 1840

\bibitem[\protect\astroncite{Hurley-Keller et~al.}{1999}]{Hurl99}
Hurley-Keller, D., Mateo, M., Grebel, E.K., 1999, ApJ, 523, L25

\bibitem[\protect\astroncite{Ibata et~al.}{1994}]{Ibata94}
Ibata, R.A., Gilmore, G., Irwin, M.J., 1994, Nature, 370, 194

\bibitem[\protect\astroncite{Ibata et~al.}{2000}]{Ibata00}
Ibata, R.A., Irwin, M.J., Lewis, G., Stolte, A., 2000, ApJ, submitted
(astro-ph/0004255)

\bibitem[\protect\astroncite{Kaluzny et~al.}{1995}]{Kal95}
Kaluzny, J., Kubiak, M., Szymanski, M., et al. 1995, A\&AS, 112, 407

\bibitem[\protect\astroncite{Karachentsev et~al.}{2000}]{Kara00}
Karachentsev, I.D., Karachentseva, V.E., Dolphin, A.E., et al., 2000,
A\&A, submitted

\bibitem[\protect\astroncite{Klessen \& Kroupa}{1998}]{Klessen98}
Klessen, R.S., Kroupa, P., 1998, ApJ, 498, 143

\bibitem[\protect\astroncite{Kleyna et~al.}{1998}]{Kleyna98}
Kleyna, J.T., Geller, M.J., Kenyon, S.J., et al., 1998, AJ, 115, 2359

\bibitem[\protect\astroncite{Kleyna et~al.}{1999}]{Kleyna99}
Kleyna, J.T., Geller, M.J., Kenyon, S.J., Kurtz, M., 1999, AJ, 117, 1275

\bibitem[\protect\astroncite{Klypin et~al.}{1999}]{Klypin99}
Klypin, A., Kravtsov, A.V., Valenzuela, O., Prada, F., 1999, ApJ, 522, 82

\bibitem[\protect\astroncite{Knapp et~al.}{1978}]{Knapp78}
Knapp, G.R., Kerr, F.J., Bowers, P.F., 1978, AJ, 83, 360

\bibitem[\protect\astroncite{Kormendy \& Bender}{1999}]{Korm99}
Kormendy, J., Bender, R., 1999, ApJ, 522, 772

\bibitem[\protect\astroncite{Kunkel}{1979}]{Kunkel79}
Kunkel, W.E., 1979, ApJ, 228, 718

\bibitem[\protect\astroncite{Lauer et~al.}{1998}]{Lauer98}
Lauer, T.R., Faber, S.M., Ajhar, C.J., et al., 1998, AJ, 116, 2263

\bibitem[\protect\astroncite{Lavery et al.}{1996}]{Lave96}
Lavery, R.J., Seitzer, P., Walker, A.R., et al., 1996, AAS, 188, 0903

\bibitem[\protect\astroncite{Layden \& Sarajedini}{2000}]{Layd00}
Layden, A.C., Sarajedini, A., 2000, AJ, 119, 1760

\bibitem[\protect\astroncite{Lee}{1995}]{Lee95}
Lee, M.G., 1995, AJ, 110, 1129

\bibitem[\protect\astroncite{Lee}{1996}]{Lee96}
Lee, M.G., 1996, AJ, 112, 1438

\bibitem[\protect\astroncite{Lee et~al.}{1993}]{Lee93}
Lee, M.G., Freedman, W., Mateo, M., et~al., 1993, AJ, 106, 1420

\bibitem[\protect\astroncite{Lee et~al.}{1999}]{Lee99}
Lee, M.G., Aparicio, A., Tikhonov, N., Byun, Y.-I., Kim, E.,
1999, AJ, 118, 853

\bibitem[\protect\astroncite{Lee \& Kim}{2000}]{Lee00}
Lee, M.G., Kim, S.C., 2000, AJ, 119, 777

\bibitem[\protect\astroncite{Lin \& Faber}{1983}]{Lin83}
Lin, D.N.C., Faber, S.M., 1983, ApJ, 266, L21

\bibitem[\protect\astroncite{Lin \& Murray}{1999}]{Lin99}
Lin, D.N.C., Murray, S.D., 1999, in
Dwarf Galaxies and Cosmology, Editions Front\`eres

\bibitem[\protect\astroncite{Mac Low \& Ferrara}{1999}]{Mac99}
Mac Low, M.-M., Ferrara, A., 1999, ApJ, 513, 142

\bibitem[\protect\astroncite{Majewski}{1999}]{Maj99}
Majewski, S.R., 1999, in Globular Clusters, eds. C. Mart\'{\i}nez Roger et al.,
Cambridge University Press

\bibitem[\protect\astroncite{Majewski et~al.}{1999}]{Maj99b}
Majewski, S.R., Ostheimer, J.C., Kunkel, W.E., et al., 1999, in
New Views of the Magellanic Clouds, IAU
Symposium 190, eds. Y.-H. Chu et al., Astronomical
Society of the Pacific

\bibitem[\protect\astroncite{Majewski et~al.}{2000}]{Maj00}
Majewski, S.R., Ostheimer, J.C., Patterson, R.J., et al., 2000, AJ, 199, 760

\bibitem[\protect\astroncite{Mart\'{\i}nez-Delgado \& Aparicio}{1998}]{Mart98}
Mart\'{\i}nez-Delgado, D., Aparicio, A., 1998, AJ, 115, 1462

\bibitem[\protect\astroncite{Massey et~al.}{1995}]{Mass95a}
Massey, P., Johnson, K.E., DeGioia-Eastwood, K., 1995a, ApJ, 454, 151

\bibitem[\protect\astroncite{Massey et~al.}{1995}]{Mass95b}
Massey, P., Lang, C.C., DeGioia-Eastwood, K., Garmany, C.D., 1995b, ApJ, 
438, 188

\bibitem[\protect\astroncite{Massey \& Hunter}{1998}]{Mass98}
Massey, P., Hunter, D., 1998, ApJ, 498, 180

\bibitem[\protect\astroncite{Mateo}{1998}]{Mateo98}
Mateo, M., 1998, ARA\&A, 36, 435

\bibitem[\protect\astroncite{Mateo et~al.}{1991}]{Mateo91}
Mateo, M., Nemec, J., Irwin, M., McMahon, R., 1991, AJ, 101, 892

\bibitem[\protect\astroncite{Mateo et~al.}{1995}]{Mateo95}
Mateo, M., Fischer, P., Krzeminski, W. 1995, AJ, 110, 2166

\bibitem[\protect\astroncite{Mighell}{1997}]{Migh97}
Mighell, K.J., 1997, AJ, 114, 1458

\bibitem[\protect\astroncite{Mighell \& Rich}{1996}]{Migh96}
Mighell, K.J., Rich, R.M., 1996, AJ, 111, 777

\bibitem[\protect\astroncite{Mighell \& Burke}{1999}]{Migh99}
Mighell, K.J., Burke, C.J., 1999, AJ, 118, 366

\bibitem[\protect\astroncite{Minniti et~al.}{1995}]{Minn95}
Minniti, D., Olszewski, E.W., Liebert, J., et al., 1995, MNRAS, 277, 1293

\bibitem[\protect\astroncite{Minniti \& Zijlstra}{1996}]{Minn96}
Minniti, D., Zijlstra, A.A., 1996, ApJ, 467, 13

\bibitem[\protect\astroncite{Minniti et~al.}{1999}]{Minn99}
Minniti, D., Zijlstra, A.A., Alonso, M.V., 1999, AJ, 117, 881

\bibitem[\protect\astroncite{Mochejska et~al.}{1998}]{Moch98}
Mochejska, B.J., Kaluzny, J.,  Krockenberger, M., 1998, AcA, 48, 455

\bibitem[\protect\astroncite{Monkiewicz et~al.}{1999}]{Monk99}
Monkiewicz, J., Mould, J.R., Gallagher, J.S., et al., 1999, PASP, 111, 1392

\bibitem[\protect\astroncite{Montegriffo et~al.}{1998}]{Monte98}
Montegriffo, P., Bellazzini, M., Ferraro, F.R., et al., 1998, MNRAS, 294, 315
 
\bibitem[\protect\astroncite{Moore et~al.}{1999}]{Moore99}
Moore, B., Ghigna, S., Governato, F., et al. 1999, ApJ, 524, L19

\bibitem[\protect\astroncite{Oh et~al.}{2000}]{Oh00}
Oh, K.S., Lin, D.N.C., Richer, H.B., 2000, ApJ, 531, 727

\bibitem[\protect\astroncite{Morrison et~al.}{2000}]{Morris00}
Morrison, H.L., Mateo, M., Olszewski, E.W.,
et al., 2000, AJ, in press (astro-ph/0001492)

\bibitem[\protect\astroncite{Olsen et~al.}{1998}]{Olsen98}
Olsen, K.A.G., Hodge, P.W., Mateo, M., et al., 1998, MNRAS, 300, 665

\bibitem[\protect\astroncite{Putman et~al.}{1998}]{Putman98}
Putman, M.E., Gibson, B.K., Staveley-Smith, L., et al., 1998,
Nature, 294, 752

\bibitem[\protect\astroncite{Richer \& McCall}{1995}]{Rich95}
Richer, M.G., McCall, M.L., 1995, ApJ, 445, 642

\bibitem[\protect\astroncite{Rocha-Pinto et~al.}{2000}]{Roch00}
Rocha-Pinto, H.J., Scalo, J., Maciel, W.J., Flynn, C., 2000, ApJ, 531, L115

\bibitem[\protect\astroncite{Rolleston et~al.}{1999}]{Roll99}
Rolleston, W.R.J., Dufton, P.L., McErlean, N.D.,
Venn, K.A., 1999, A\&A, 348, 728

\bibitem[\protect\astroncite{Sage et~al.}{1998}]{Sage98}
Sage, L.J., Welch, G.A., Mitchell, G.F., 1998, ApJ, 507, 726

\bibitem[\protect\astroncite{Sakai et~al.}{1999}]{Sakai99}
Sakai, S., Madore, B.F., Freedman, W.L., 1999, ApJ, 511, 671

\bibitem[\protect\astroncite{Sarajedini et~al.}{1998}]{Sara98}
Sarajedini, A., Geisler, D., Harding, P., Schommer, R., 1998, ApJ, 508, L37

\bibitem[\protect\astroncite{Saviane et~al.}{1996}]{Savi96}
Saviane, I., Held, E.V., Piotto, G., 1996, A\&A, 315, 40

\bibitem[\protect\astroncite{Saviane et~al.}{2000}]{Savi00}
Saviane, I., Held, E.V., Bertelli, G., 2000, A\&A, 355, 56

\bibitem[\protect\astroncite{Searle \& Zinn}{1978}]{SZ78}
Searle, L., Zinn, R., 1978, ApJ, 225, 357

\bibitem[\protect\astroncite{Shetrone et~al.}{1998}]{Shet98}
Shetrone, M.D., Bolte, M., Stetson, P.B., 1998, AJ, 115, 1888

\bibitem[\protect\astroncite{Smecker-Hane et~al.}{1996}]{Smec96}
Smecker-Hane, T.A., Stetson, P.B., Hesser, J.E., Vandenberg, D.A.,
1996, in From Stars to Galaxies, ed. C.\ Leitherer et al.,
Astronomical Society of the Pacific

\bibitem[\protect\astroncite{Smecker-Hane et~al.}{2000}]{Smec00}
Smecker-Hane, T.A., Mandushev, G.I., Hesser, J.E., et al.,
2000, in Spectro-Photometric 
Dating of Stars and Galaxies, eds. I. Hubeny, S. Heap, \& R. Cornett, 
Astronomical Society of the Pacific

\bibitem[\protect\astroncite{Smith et~al.}{2000}]{Smith00}
Smith, G.H., Holtzman, J.A., Grillmair, C.J., 2000, in prep.

\bibitem[\protect\astroncite{Stetson et~al.}{1998}]{Stet98}
Stetson, P.B., Hesser, J.E., Smecker-Hane, T.A., 1998, PASP, 110, 533

\bibitem[\protect\astroncite{Tikhonov}{1999}]{Tikh99}
Tikhonov, N.A., 1999, in The Stellar Content of the Local Group,
IAU Symposium 193, eds. P. Whitelock \& R. Cannon, Astronomical
Society of the Pacific 

\bibitem[\protect\astroncite{Tolstoy et~al.}{1998}]{Tols98}
Tolstoy, E., Gallagher, J.S., Cole, A.A., et al., 1998, AJ, 116, 1244

\bibitem[\protect\astroncite{Tolstoy et~al.}{2000}]{Tols00}
Tolstoy, E., Gallagher, J.S., Greggio, L., et al., 2000, The ESO Messenger,
99, 16 

\bibitem[\protect\astroncite{Trentham}{1998}]{Trent98}
Trentham, N., 1998, MNRAS, 294, 193
 
\bibitem[\protect\astroncite{Tully \& Fisher}{1987}]{Tully87}
Tully, R.B., Fisher, J.R., 1987, Nearby Galaxies Atlas,
Cambridge University Press

\bibitem[\protect\astroncite{Unavane et~al.}{1996}]{Una96}
Unavane, M., Wyse, R.F.G., Gilmore, G., 1996, MNRAS, 278, 727

\bibitem[\protect\astroncite{van den Bergh}{1994}]{vdB94}
van den Bergh, S., 1994, ApJ, 428, 617 

\bibitem[\protect\astroncite{van den Bergh}{1999a}]{vdB99a}
van den Bergh, S., 1999a, ApJ, 517, L97

\bibitem[\protect\astroncite{van den Bergh}{1999b}]{vdB99b}
van den Bergh, S., 1999b, A\&ARv, 9, 273

\bibitem[\protect\astroncite{van den Bergh}{2000a}]{vdB00}
van den Bergh, S., 2000a, in The Local Group,
Cambridge University Press

\bibitem[\protect\astroncite{van den Bergh}{2000b}]{vdB00b}
van den Bergh, S., 2000b, PASP, in press (astro-ph/0001040)

\bibitem[\protect\astroncite{van der Marel et~al.}{1997}]{Marel97}
van der Marel, R.P., de Zeeuw, P.T., Rix, H.-W., Quinlan, G.D., 1997, 
Nature, 385, 610

\bibitem[\protect\astroncite{Weinberg}{2000}]{Wein00}
Weinberg, M.D., 2000, ApJ, 532, 922

\bibitem[\protect\astroncite{Westerlund}{1997}]{Wester97}
Westerlund, B.E., 1997, The Magellanic Clouds, Cambridge University Press

\bibitem[\protect\astroncite{Wilcots \& Miller}{1998}]{Wil98}
Wilcots, E.M., Miller, B.W., 1998, AJ, 116, 2363

\bibitem[\protect\astroncite{Wilkinson \& Evans}{1999}]{Wilk99}
Wilkinson, M.I., Evans, N.W., 1999, MNRAS, 310, 645
 
\bibitem[\protect\astroncite{Whiting et~al.}{1999}]{Whit99}
Whiting, A.B., Hau, G.K.T., Irwin, M. 1999, AJ, 118, 2767

\bibitem[\protect\astroncite{Young \& Lo}{1996}]{Young96}
Young, L.M., Lo, K.Y., 1996, ApJ, 464, L59

\bibitem[\protect\astroncite{Young \& Lo}{1997a}]{Young97a}
Young, L.M., Lo, K.Y., 1997a, ApJ, 476, 127

\bibitem[\protect\astroncite{Young \& Lo}{1997b}]{Young97b}
Young, L.M., Lo, K.Y., 1997b, ApJ, 490, 710

\bibitem[\protect\astroncite{Young}{1999}]{Young99}
Young, L.M., 1999, AJ, 117, 1758

\bibitem[\protect\astroncite{Young}{2000}]{Young00}
Young, L.M., 2000, AJ, 119, 188

\bibitem[\protect\astroncite{Zaritsky et~al.}{2000}]{Zar00}
Zaritsky, D., Harris, J., Grebel, E.K., 
Thompson, I.B., 2000, ApJ, 534, L53

\end{thebibliography}
\end{document}